\documentclass[conference,a4paper]{IEEEtran}

\usepackage[utf8]{inputenc} 
\usepackage[T1]{fontenc}
\usepackage{url}              
\usepackage{cite}             

\usepackage[cmex10]{amsmath}  

\DeclareMathOperator*{\argmax}{arg\,max}

\usepackage{mleftright}       
\mleftright                   

\usepackage{graphicx}         
\usepackage{booktabs}         

\usepackage{soul}

\usepackage{xcolor}

\allowdisplaybreaks[2]
\graphicspath{{figures/}}

\usepackage{amsthm}
\usepackage{amsmath}
\usepackage{amsfonts}
\usepackage{cancel}

\newtheorem*{theorem}{Theorem}

\usepackage{subcaption}

\usepackage{cleveref}
\crefname{equation}{Eq}{Eqs} 
\crefrangelabelformat{equation}{(#3#1#4--#5#2#6)}

\usepackage{float}

\usepackage{tikz}
\usetikzlibrary{backgrounds, calc, chains, fit, matrix, positioning, shadows, shapes, intersections, circuits.ee}
\tikzset{input/.style={}}
\tikzset{output/.style={}}
\tikzset{op/.style={circle, draw, fill=black!10, minimum size=2.5ex, inner sep=0ex}}
\tikzset{filter/.style={rectangle, draw, thick, fill=black!10, minimum size=3.5ex, inner sep=1ex}}
\tikzset{nn/.style={trapezium, trapezium angle=80, draw, thick, fill=black!10, inner sep=1ex}}
\tikzset{branch/.style={circle, draw, thick, fill=black, minimum size=.5ex, inner sep=0ex}}
\tikzset{tensor/.style={rectangle, draw, fill=white, minimum size=2em, double copy shadow={shadow xshift=.5ex,shadow yshift=-.5ex}}}
\tikzset{rounded/.style={rounded rectangle, draw, thick, fill=black!10, minimum size=3.5ex, inner xsep=1ex}}
\tikzset{image/.style={rectangle, draw, fill=white, minimum size=2em}}
\tikzset{>=direction ee}
\tikzset{/tikz/thin/.style={line width=.9pt}}
\tikzset{/tikz/thick/.style={line width=1.4pt}}
\tikzset{every path/.style={thin}}

\usepackage{pgfplots}
\pgfplotsset{compat=1.14}
\pgfplotsset{every axis/.append style={enlargelimits={abs=3pt},grid,axis lines=left}}
\pgfplotsset{every axis plot/.append style={thick,mark size=1.5pt,line join=bevel,mark options={solid}}}
\pgfplotsset{label style={font=\small}}
\pgfplotsset{tick label style={font=\footnotesize}}
\pgfplotsset{grid style={color=black!10}}
\pgfplotsset{legend style={draw=none,opacity=.85,font=\footnotesize,cells={anchor=west,opacity=1}}}
\pgfplotsset{every non boxed x axis/.style={xtick align=center,shorten <=-.5\pgflinewidth}}
\pgfplotsset{every non boxed y axis/.style={ytick align=center,shorten <=-.5\pgflinewidth}}
\pgfplotsset{every non boxed z axis/.style={ztick align=center,shorten <=-.5\pgflinewidth}}
\pgfplotsset{/pgf/number format/1000 sep={\,}}

\IEEEoverridecommandlockouts

\begin{document}

\title{Learned Wyner--Ziv Compressors Recover Binning\thanks{This work is supported in part by NYU Wireless and Google.}} 

\author{%
  \IEEEauthorblockN{Anonymous Authors}
  \IEEEauthorblockA{%
    Please do NOT provide authors' names and affiliations\\
    in the paper submitted for review, but keep this placeholder.\\
    ISIT23 follows a \textbf{double-blind reviewing policy}.}
}

\author{%
  \IEEEauthorblockN{Ezgi~{\"O}zyılkan}
  \IEEEauthorblockA{Dept.~of Electrical and Computer Engineering \\
  New York University \\
  NY 11201, USA \\
\texttt{ezgi.ozyilkan@nyu.edu}}
  \and
  \IEEEauthorblockN{Johannes~Ballé}
  \IEEEauthorblockA{Google Research \\ 
                    New York \\
                    NY 10011, USA \\
                   \texttt{jballe@google.com}}
\and 
\IEEEauthorblockN{Elza Erkip}
  \IEEEauthorblockA{Dept.~of Electrical and Computer Engineering \\
  New York University \\
  NY 11201, USA \\
\texttt{elza@nyu.edu}}
}

\maketitle

\begin{abstract}
  We consider lossy compression of an information source when the decoder has lossless access to a correlated one. This setup, also known as the \emph{Wyner--Ziv} problem, is a special case of distributed source coding. To this day, real-world applications of this problem have neither been fully developed nor heavily investigated. We propose a data-driven method based on machine learning that leverages the universal function approximation capability of artificial neural networks. We find that our neural network-based compression scheme re-discovers some principles of the optimum theoretical solution of the Wyner--Ziv setup, such as binning in the source space as well as linear decoder behavior within each quantization index, for the quadratic-Gaussian case. These behaviors emerge although no structure exploiting knowledge of the source distributions was imposed. Binning is a widely used tool in information theoretic proofs and methods, and to our knowledge, this is the first time it has been explicitly observed to emerge from data-driven learning.
\end{abstract}

\section{Introduction}
\label{sec:intro}

Consider a distributed sensor network consisting of individual cameras that independently capture images at different locations across the same city. Suppose that each sensor node compresses and transmits its highly correlated image to a joint central processing unit that reproduces a unified visual map of the city, by fusing the information collected by all of the nodes. If the sensors could directly communicate with each other in a cooperative manner, they could avoid some degree of redundancy by transmitting less correlated information. However, direct communication between nodes is often infeasible.

Given that, what is the best strategy to exploit the correlation between sensor data? Slepian and Wolf~\cite{Slepian:IT:73} (SW) proved a remarkable and well-known information theoretic result that the distributed compression is asymptotically as efficient as the joint one, if the joint distribution statistics are known and compression is lossless. Their proof invokes \emph{random binning} arguments and is non-constructive. Establishing a practical framework building onto these concepts is a challenging open problem to this day.

\begin{figure}
\centering
\begin{subfigure}[b]{\columnwidth}
   \includegraphics[width=1\linewidth]{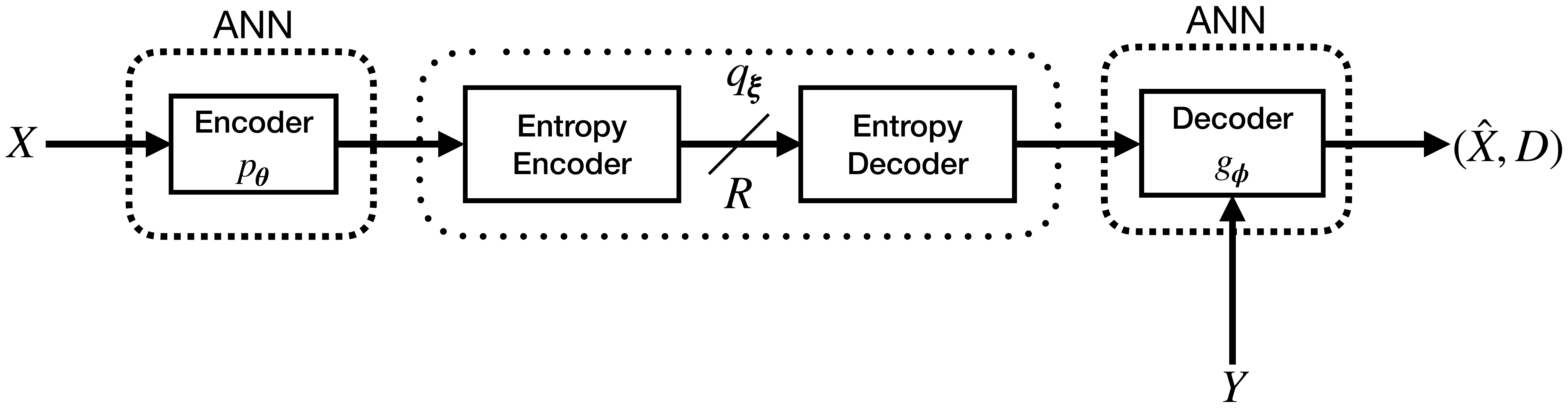}
   \caption{}
   \label{fig:sys_marginal} 
\end{subfigure}

\begin{subfigure}[b]{\columnwidth}
   \includegraphics[width=1\linewidth]{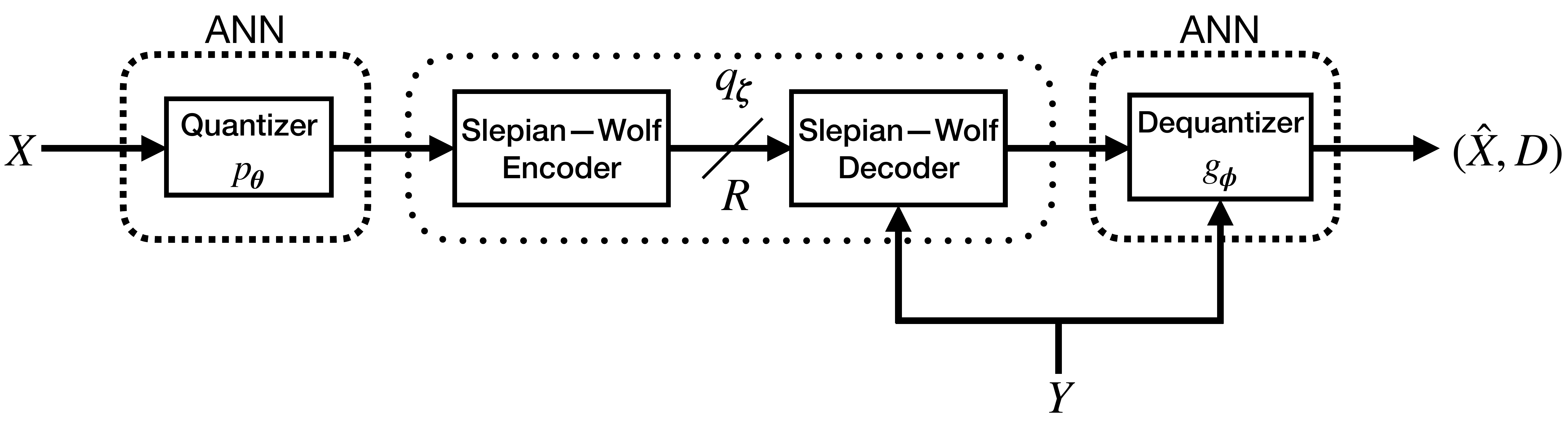}
   \caption{}
   \label{fig:sys_conditional}
\end{subfigure}

\caption{The two lossy compression systems that we consider: learned compressor using a classic entropy coder (a) and learned quantizer and dequantizer, using an ideal Slepian--Wolf coder (b).}
\label{fig:sys}
\end{figure} 

Here, we investigate the setup characterized by Wyner and Ziv~\cite{Wyner:IT:76} (WZ), which is both more general than SW as it encompasses lossy compression, and a simpler special case, as it assumes the decoder has access to a correlated source, the \emph{side information}, losslessly. For WZ coding, there has been vast prior work considering synthetic setups and specific correlation patterns. Zamir \emph{et al.}~\cite{zamir_ITW, zamir_TIT} outlined the asymptotically optimal constructive mechanisms using nested linear and lattice codes for binary and Gaussian sources, respectively. Since then, the constructive and non-asymptotic research effort has been spearheaded by distributed source coding using syndromes (DISCUS)~\cite{DISCUS}, which formulated the WZ setup as a dual quantizer-channel coding problem. In a nutshell,
the source is first quantized in a suitable manner according to its marginal density. Next, the quantization codebook space is partitioned into \emph{cosets}, according to the virtual channel arising between the side information and the quantized source. Instead of sending the quantization index to the decoder, the encoder only sends the index of the coset containing the quantized codeword, which results in further rate reduction. The decoder can then disambiguate the coset index with the help of the side information, and recovers the quantization index. Finally, it estimates the source using the deduced quantization index and the side information, according to the distortion criterion. Note that such a systematic partitioning of the quantized source space with cosets effectively mimics the random binning procedure in the proofs of the SW and also of the WZ theorem that consider the asymptotic regime. The complex interaction between the quantization, channel coding and estimation parts was also highlighted in competitive practical code design frameworks proposed in~\cite{swc_nq, tcq_ldpc}. These methods achieve performances close to the theoretical bound, but are only applicable for Gaussian sources.

We propose to leverage the universal function approximation capability of artificial neural networks (ANNs)\cite{Leshno1993, hornik_et_al} and machine learning techniques to find constructive solutions for the non-asymptotic regime. More specifically, we consider the \emph{one-shot} case, i.e., compressing each source realization one at a time, similarly to popular ANN-based compressors~\cite{Balle2017, balle_journal}. We provide two distinct solutions to the WZ problem, where we either handle the quantization and binning parts jointly (Fig.~\ref{fig:sys_marginal}) or take a two-step approach by having a learned quantizer that is coupled with an ideal SW coder (Fig.~\ref{fig:sys_conditional}). We defer discussion of the entropy coders, shown in Fig.~\ref{fig:sys}, to later sections. 

In order to establish the training objectives for these solutions, aiming for optimality, we minimize upper bounds on mutual information. These are expressed through one of the two probabilistic models utilizing ANNs (Section~\ref{sec:method}). Next, we explain how each probabilistic model is interpretable as one of the operational schemes shown in Fig.~\ref{fig:sys} (Sections \ref{sec:marginal} and \ref{sec:conditional}, respectively). We discuss empirical results and connections to related work in Section~\ref{sec:discussion}.

\section{Estimating neural upper bounds on Wyner--Ziv}
\label{sec:method}
Since our choice of objective functions is inspired by the rate--distortion function of the case where side information is only available at the decoder, we briefly recap the WZ theorem and the accompanying information theoretic concepts. For the complete proof, refer to the original paper~\cite{Wyner:IT:76} and to~\cite{network_info_theo}.
\begin{theorem} \label{theo:WZ}(Wyner--Ziv Theorem [1976]) Let $(X,Y)$ be correlated sources, drawn i.i.d. $\sim p(x,y)$, and let $d(x, \hat{x})$ be a single-letter distortion measure. The rate--distortion function for $X$ with side information $Y$
available (non-causally) at the decoder side is as follows:
\begin{equation} \label{eq:WZ}
    R_{\text{WZ}}(D) = \min (I(X;U) - I(Y;U)),
\end{equation}
where the minimization operation is over
all conditional probability distribution functions $p(u \vert  x)$,
and all functions $g(u,y)$ such that  $ \mathbb{E}_{p(x,y)p(u\vert x)} d(x, g(u,y) ) \le D$. \end{theorem}The achievability part of the WZ theorem invokes the covering lemma, resulting in the rate of $I(X;U)$, followed by a random binning argument based on joint typicality, which yields the rate discount of $I(Y;U)$ in Eq.~\eqref{eq:WZ}~\cite{elements_of_information_theory}. This achievability, which is shown to be tight, assumes a Markov chain constraint $U-X-Y$. 

Assuming further that the encoder in the achievability proof is represented by a probability model $p_{\boldsymbol{\theta}}(u\vert x)$ with parameters $\boldsymbol \theta$, the mutual information in Eq.~\eqref{eq:WZ} can be written as:
\begin{align}
    I(X;U) - I(Y;U) &= I(X;U \vert Y) \nonumber \\
     &= \mathbb{E}_{\substack{p(x,y)\\p_{\boldsymbol{\theta}}(u\vert x)}} \log \frac{p_{\boldsymbol{\theta}}(u \vert  x)}{\cancel{p(u)}} \cdot \frac{\cancel{p(u)}}{p (u \vert y)}.
\end{align}
We will use the probabilistic model $p_{\boldsymbol{\theta}}(u\vert x)$ to facilitate the learning procedure of an encoder, and we set our encoder output as $u = \argmax_v p_{\boldsymbol{\theta}}(v\vert x)$. To consider a practical source coding setting, we also have $U$ as discrete. For our objective functions, we choose one of two variational upper bounds:
\begin{align}
    I(X;U \vert Y) &\leq \mathbb{E}_{\substack{p(x,y)\\p_{\boldsymbol{\theta}}(u\vert x)}} \log\frac{p_{\boldsymbol{\theta}}(u \vert  x )}{q_{\boldsymbol{\xi}} (u)}, \label{eq:upper_bound_marg} \\
    I(X;U \vert Y) &\leq \mathbb{E}_{\substack{p(x,y)\\p_{\boldsymbol{\theta}}(u\vert x)}} \log\frac{p_{\boldsymbol{\theta}}(u \vert  x )}{q_{\boldsymbol{\zeta}} (u \vert  y)}.  \label{eq:upper_bound_cond}
\end{align}
Here, $q_{\boldsymbol{\xi}}(u)$ and $q_{\boldsymbol{\zeta}}(u \vert y)$ (with parameters $\boldsymbol \xi$ and $\boldsymbol \zeta$, respectively), are two different models of the distribution $p( u \vert y)$, which is generally not known in closed form. We will discuss the operational meaning of these two variants in Sections~\ref{sec:marginal} and \ref{sec:conditional}. The upper bounds in Eqs.~\eqref{eq:upper_bound_marg} and \eqref{eq:upper_bound_cond} follow from cross-entropy~\cite{kullback} being larger or equal to entropy~\cite{elements_of_information_theory}.

We define all probabilistic models $p_{\boldsymbol{\theta}}(u \vert x)$, $q_{\boldsymbol{\xi}}(u)$ and $q_{\boldsymbol{\zeta}}(u \vert y)$, as discrete distributions with probabilities
\begin{equation} \label{eq:categorical_distribution}
    P_k = \frac{\exp \alpha_k}{\sum_{i=1}^K\exp \alpha_i },
\end{equation}
for $k \in \{1, \dots, K\}$, where $K$ is a model parameter. The unnormalized log-probabilities (\emph{logits}) $\alpha_k$ are computed by ANNs as functions of the conditioning variable (i.e., $x$ for $p_{\boldsymbol{\theta}}(u \vert x)$ and $y$ for $q_{\boldsymbol{\zeta}}(u \vert y)$), where the parameters represent the ANN weights, or treated as learnable parameters directly (for $q_{\boldsymbol{\xi}}(u)\, $). This choice keeps the parametric families as general as possible and does not unnecessarily impose any structure. Specifically, this allows the model $p_{\boldsymbol{\theta}}(u \vert x)$ to learn, if needed, quantization schemes that involve discontiguous bins, akin to the \emph{random binning} operation in the achievability part of the WZ theorem~\cite{Wyner:IT:76}, and resembling the systematic partitioning of the quantized source space with cosets in DISCUS~\cite{DISCUS}.

Next, we relax the constrained formulation of the WZ theorem to an unconstrained one using a Lagrange multiplier. Building on the upper bounds developed in Eqs.~\eqref{eq:upper_bound_marg} and \eqref{eq:upper_bound_cond}, this yields either a marginal or a conditional loss function:
\begin{align} \label{eq:proposed_loss_marginal}
     L_\mathrm{m}(\boldsymbol{\theta}, \boldsymbol{\phi}, \boldsymbol{\xi}) &= \mathbb{E} \Big[\log \frac{p_{\boldsymbol{\theta}}(u \vert x)}{q_{\boldsymbol{\xi}}(u)} + \lambda  \cdot  d(x, g_{\boldsymbol{\phi}}(u, y))\Big], \\ \label{eq:proposed_loss_conditional}
     L_\mathrm{c}(\boldsymbol{\theta}, \boldsymbol{\phi}, \boldsymbol{\boldsymbol{\boldsymbol{\zeta}}}) &= \mathbb{E} \Big[\log \frac{p_{\boldsymbol{\theta}}(u \vert x)}{q_{\boldsymbol{\zeta}}(u \vert y)} + \lambda \cdot d(x, g_{\boldsymbol{\phi}}(u, y))\Big],
\end{align}
where $\{\boldsymbol{\theta}, \boldsymbol{\phi}, \boldsymbol{\xi}, \boldsymbol{\zeta}\}$ are optimization parameters, and $g_{\boldsymbol{\phi}}(u,y)$ is the decoding function, also represented by an ANN with parameters $\boldsymbol{\phi}$, which outputs the reconstruction $\hat{x}=g_{\boldsymbol{\phi}}(u, y)$. We dropped the subscript from the expectation operator for brevity. The optimized $p_{\boldsymbol{\theta}}(u\vert x)$ and $g_{\boldsymbol{\phi}}(u,y)$ models yield the ANN-based encoder--decoder and quantizer--dequantizer components, respectively, considering system models in Figs.~\ref{fig:sys_marginal} and \ref{fig:sys_conditional}. The learnable parameters can be jointly optimized with stochastic gradient descent (SGD) since the loss functions are differentiable with respect to them. We can compute the gradients using automatic differentiation methods, as implemented in deep learning frameworks such as JAX~\cite{jax}. By varying the trade-off parameter $\lambda$, we obtain different points in the achievable rate--distortion region.
SGD replaces the expectations in the loss functions by averages over batches of samples $B$, and exchanges the order of differentiation and summation, due to linearity. For a sample loss $\ell_{\boldsymbol{\theta}}(x,y)$ with parameters $\boldsymbol \theta$ (represented as one of the sample loss functions inside the brackets in Eqs.~\eqref{eq:proposed_loss_marginal} or \eqref{eq:proposed_loss_conditional}) we approximate:
\begin{align}
    \frac{\partial }{\partial \boldsymbol{\theta} }\mathbb{E} [\;  \ell_{\boldsymbol{\theta}}(x,y)) \; ] \approx \frac{1}{|B|} \sum_{(x, y)\in B} \frac{\partial \ell_{\boldsymbol{\theta}}(x,y)}{\partial \boldsymbol{\theta}} \; . 
\end{align} 

This requires drawing samples $u$ from the model $p_{\boldsymbol{\theta}}(u\vert x)$ throughout training. A well-known technique, termed Gumbel-max trick originally proposed in~\cite{gumbel_org}, offers a way to draw samples from any discrete distribution. The trick proposes to draw a sample from a distribution of $K$ states (as in Eq.~\eqref{eq:categorical_distribution}):
\begin{equation} \label{eq:gumbel_max}
    \argmax _{k \in \{1, \dots, K \} }\{\alpha_{k} + G_{k} \},
\end{equation}
where $G_{k}$ are i.i.d. samples from a standard Gumbel distribution. 
Observing that the derivative of the $\argmax$ operator in Eq.~\eqref{eq:gumbel_max} is equal to zero everywhere except at the boundaries of state changes, we note that we instead need a \emph{continuous relaxation} of this operator during training in order to carry out SGD. The Concrete distribution, proposed in~\cite{concrete}, establishes a way to do that. Instead of drawing discrete (hard) samples, this approach allows us to obtain soft samples, a vector of length $K$ where the mass is spread out across multiple states rather than being concentrated in one. The index $k \in \{1, \dots, K\}$ of such a soft sample is defined using a \emph{softmax} function as, 
\begin{equation}
    U_k = \frac{\exp((\alpha_{k}+G_k)\; / \; t)}{\sum_{i=1}^K\exp((\alpha_{i}+G_i)\; /\; t)} \; ,\label{eq:softmax}
\end{equation}
where $t$ is a temperature parameter that dictates the amount of relaxation. Note that in the limit of $t \rightarrow 0^{+}$, the soft samples converge to their hard counterparts, which also means that the Concrete distribution itself converges to a discrete one. To match the distribution of the $u$ samples, we also choose $q_{\boldsymbol{\xi}}(u)$ and $q_{\boldsymbol{\zeta}}(u \vert y)$ as Concrete during training.

\subsection{Evaluation and experimental setup}
The WZ formula in Eq.~\eqref{eq:WZ} has a closed-form expression only in a few special cases\footnote{Although the Blahut--Arimoto (BA) algorithm~\cite{blahut, arimoto} can be used to compute the rate--distortion function, it is noted that BA provides inaccurate estimates in continuous settings and fails to scale with high-dimensional datasets~\cite{neural_RD, empirical_sandwich}.}. To evaluate how close our neural bounds get to the known rate--distortion function, we consider the following correlation model: let $X$ and $Y$ be correlated, zero mean and stationary Gaussian memoryless sources, and let the distortion metric be mean-squared error. Then, the WZ rate--distortion function is
\begin{equation} \label{eq:WZ_CD}
    R_{\text{WZ}}(D) =  \frac{1}{2}\log\left( \frac{\sigma_{x \vert y}^{2}}{D} \right), \; \; 0 \leq D \leq \sigma_{x \vert y}^{2},
\end{equation}
where $ \sigma_{x \vert y}^{2}$ denotes the conditional variance of $X$ given $Y$. For $X=Y+N$, where $N \sim \mathrm{N}(0, \sigma_n^2)$, which is considered throughout the paper except Fig.~\ref{fig:RD_diff}, we have $\sigma_{x \vert y}^2 = \sigma_{n}^2$. The rate--distortion function for $Y=X+N$, considered in Fig.~\ref{fig:RD_diff}, can also be derived similarly.

Note that in spite of considering Gaussian sources, we do \emph{not} make any assumptions on the distribution of information sources in our formulations of the models. The parameters $\{\boldsymbol{\theta}, \boldsymbol{\phi}, \boldsymbol{\xi}, \boldsymbol{\zeta}\}$ are learned solely in a data-driven way from realizations of the sources, through the proposed loss functions in Eqs.~\eqref{eq:proposed_loss_marginal} and \eqref{eq:proposed_loss_conditional}.

For the conditional probabilistic models and the decoding function, we employ ANNs of three dense layers, with 100 units each (excluding the last one), and leaky rectified linear units as activation functions (again, excluding the last) for each of the layers. In our experiments, we found that larger networks or different activation functions did not improve the results. The decoding function receives a concatenated vector of both its inputs, $u$ and $y$. We use Adam~\cite{adam}, a popular variant of SGD, and conduct our experiments using the JAX~\cite{jax} framework. For evaluation, we switch from Concrete distributions back to their discrete counterparts, and use a deterministic encoding function that is equal to the mode of $p_{\boldsymbol{\theta}}(u \vert x)$, rather than sampling from it. We obtain all empirical estimates of rate and distortion by averaging over $2^{20}$ source realizations.

\section{Operational meaning and evaluation of $L_\mathrm{m}$} \label{sec:marginal}

\begin{figure*}[t]
\centering
\begin{subfigure}{.5\textwidth}
  \centering
  \includegraphics[width=0.85\linewidth]{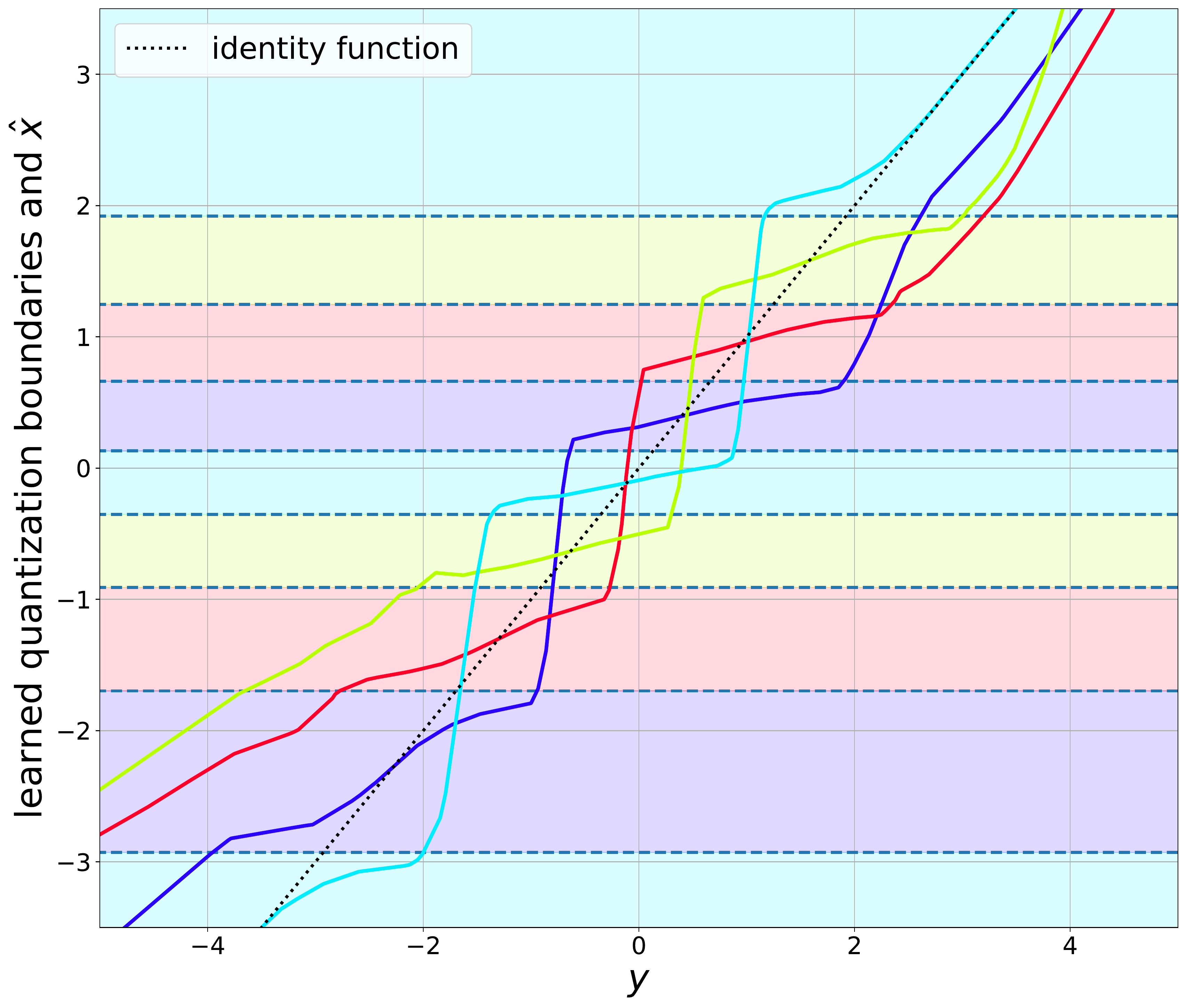}
  \caption{$X=Y+N$ with $Y \sim N(0,1)$ and $\mathrm{N} \sim N(0,10^{-1})$.}
  \label{fig:visualization_1}
\end{subfigure}%
\hfill
\begin{subfigure}{.5\textwidth}
  \centering
  \includegraphics[width=0.85\linewidth]{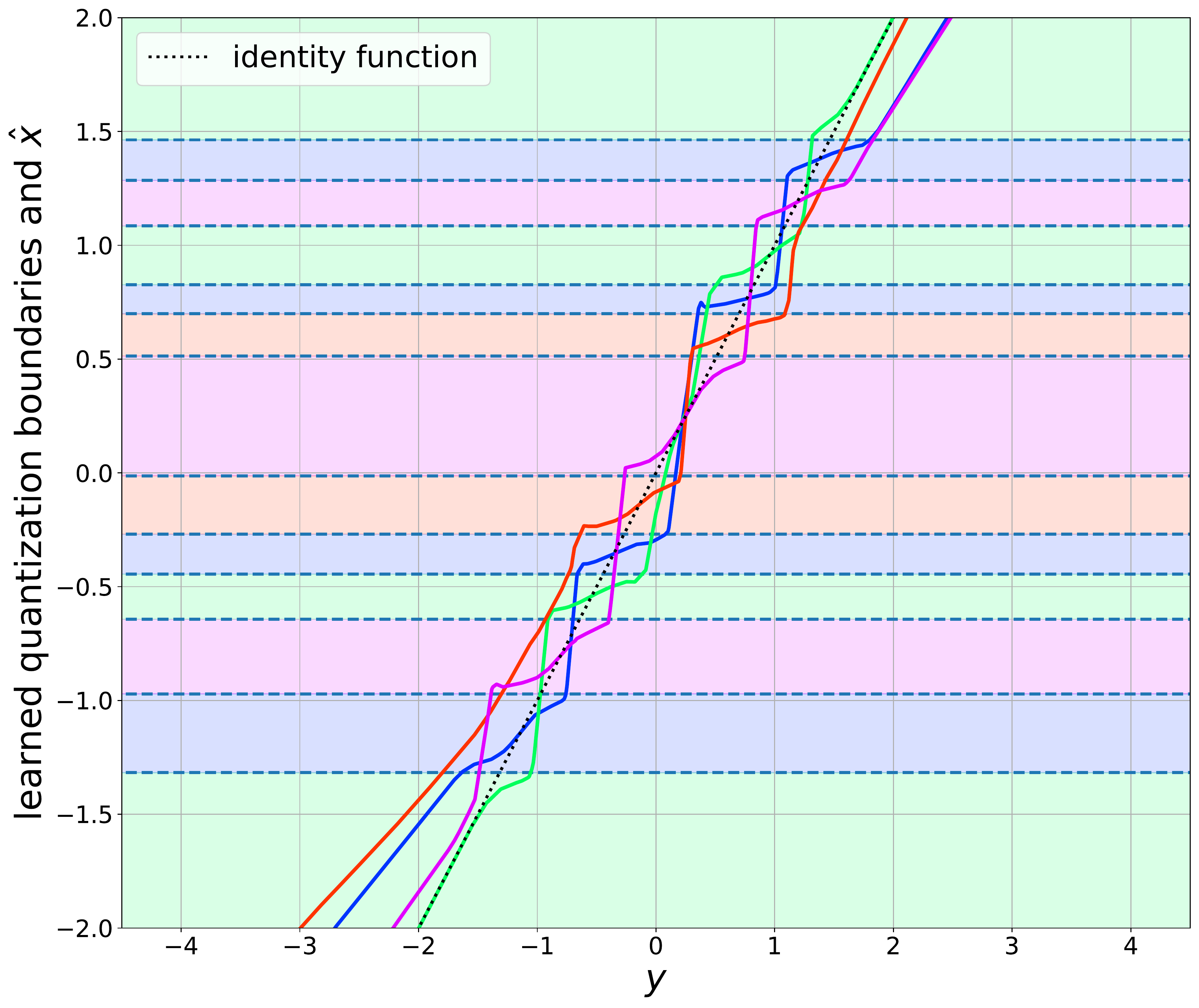}
  \caption{$X=Y+N$ with $Y \sim N(0,1)$ and $\mathrm{N} \sim N(0,10^{-2})$.}
  \label{fig:visualization_2}
\end{subfigure}
\caption{Visualizations (best viewed in color) of the learned deterministic encoder $u = \argmax_v p_{\boldsymbol{\theta}}(v\vert x)$ and decoder $\hat{x} = g_{\boldsymbol{\phi}}(u,y)$ of the marginal formulation (Eq.~\eqref{eq:proposed_loss_marginal}), for the quadratic--Gaussian WZ setup. The dashed horizontal lines are quantization boundaries, and the colors between boundaries represent unique values of $u$. We depict the decoding function as separate plots for each value of $u$, using the same color assignment. Rate--distortion performance of the model depicted in (a) is provided in Fig.~\ref{fig:RD_var0.1}. The model in (b) achieves $-23.07$ dB at $1.94$ bits.}
\label{fig:visualizations}
\end{figure*}

We first consider the system model in Fig.~\ref{fig:sys_marginal}. Note that the upper bound in Eq.~\eqref{eq:upper_bound_marg} corresponds to the rate of a system employing a one-shot encoder and an entropy code which asymptotically achieves a rate equal to the cross-entropy $\mathbb{E}_{x}\big[\mathbb{E}_{u \sim p_{\boldsymbol{\theta}}(u \vert x)}[-\log q_{\boldsymbol{\xi}}(u)] \big]$. Therefore, by minimizing $L_\mathrm{m}$ in Eq.~\eqref{eq:proposed_loss_marginal}, we optimize such an operational scheme in an end-to-end fashion.

In Figs.~\ref{fig:visualization_1} and \ref{fig:visualization_2}, we visualize the learned compressors obtained with this formulation. We remark that the learned compressors exhibit periodic grouping, binning-like behavior with respect to the source space, although no explicit structure was imposed onto the model architecture. Color coding of the bin indices reveals discontiguous quantization bins. This demonstrates that ANN-based methods are indeed capable of recovering very similar solutions to some of the handcrafted frameworks proposed for the WZ problem, such as DISCUS~\cite{DISCUS}. Note that this behavior is also analogous to the random binning procedure in the achievability part of the WZ theorem~\cite{Wyner:IT:76}.

The figure also shows that the learned compressors exhibit optimal decoder behavior within each quantization index. In the given setup, the optimal decoder disambiguates the quantization index from the received bin index $u$, and reconstructs the source as~\cite{zamir},
\begin{equation} \label{eq:opt_decoder}
    \hat{x} = (1-\beta) \cdot y + \beta \cdot M(u), \;  \text{where} \; \beta \propto \sigma_{n}^2 \: ,
\end{equation}
where $M(\cdot)$ denotes the disambiguation procedure. The slopes of the learned curves are also sensitive to $\sigma_{n}^2$, as is evident from comparing both panels of Fig.~\ref{fig:visualizations}.

We explain the behavior of the learned encoder and decoder as follows. The encoder quantizes the source and subsequently bins the quantization index using the learned joint statistics of $Q(X)$ and $Y$, where $Q(\cdot)$ refers to the quantization, yielding $u$. Note that the encoder does \emph{not} explicitly have access to the realization $Y=y$. The decoder then disambiguates the received bin index and deduces the quantization index, with the help of the side information. It subsequently estimates the source as $\hat{x}$, yielding the linear decoding functions within each quantization index with respect to the matching curve shown in Fig.~\ref{fig:visualizations}.  Observe that the corresponding decoding functions in matching quantization indices exhibit kinks close to the boundaries. This demonstrates that the model tries to adopt linear functions, as is the optimal decoder behavior in Eq.~\eqref{eq:opt_decoder}.

As seen in Fig.~\ref{fig:RD_var0.1}, our learned compressor yields a better performance compared to the point-to-point rate--distortion function. We argue that this is mainly due to the learned binning behavior, resulting in rate reduction. However, the compressor does not reach the asymptotic WZ rate--distortion bound provided in Eq.~\eqref{eq:WZ_CD}. In the figure, $1.53$ dB refers to the mean-squared error gap that the entropy-constrained scalar (one-shot) lattice quantizer is subjected to in a high-rate regime \cite{quantization}, due to space-filling loss (also known as cubic loss~\cite{cubic_loss}). As the ANN model compresses and consecutively bins each scalar input one by one, it is subjected both to the space-filling loss~\cite{high_resolution_quantization} during the quantization step, as well as to the loss coming from binning non-uniformly distributed quantization indices. The achievability part of the WZ theorem, by comparison, considers binning of long sequences. This type of \emph{compress--bin}~\cite{network_info_theo} is much more efficient than the one-shot case we consider, as it exploits the correlated side information in a better way. Note that visualizing the system behavior as in Fig.~\ref{fig:visualizations} is difficult for more than one-dimensional $x$ and $y$ (i.e., for larger block lengths). 

\section{Operational meaning and evaluation of $L_\mathrm{c}$}
\label{sec:conditional}
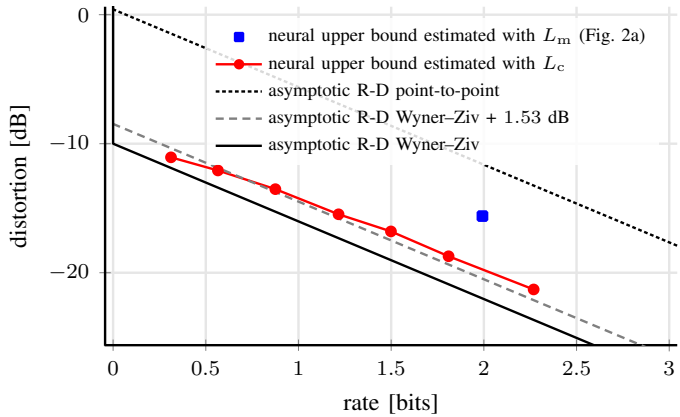
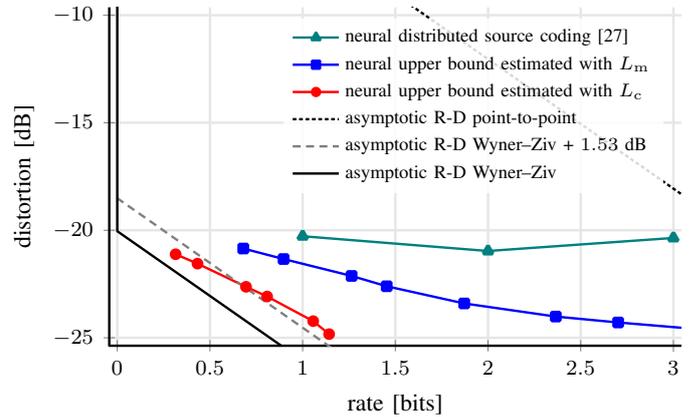
\begin{figure*}[t]
\begin{subfigure}{\columnwidth}
    \raggedleft
    \begin{tikzpicture}[trim axis right]
    \begin{axis}[
      height=.19\textheight,
      width=.85\linewidth,
      scale only axis,
      xlabel={rate [bits]},
      ylabel={distortion [dB]},
      xmin=0.,
      xmax=3.,
      ymin=-25.,
      ymax=0.,
      legend pos=north east,
      legend style={font=\scriptsize},
      ]
       \addplot[color=blue,mark=square*,only marks] table {
        1.992392 -15.610814094543457
      };
      \addplot[color=red,mark=*] table {
        0.31238356 -11.05667233467102
        0.5654873 -12.075015306472778
        0.87530893 -13.522629737854004
        1.2161504 -15.470525026321411
        1.4988127 -16.797856092453003
        1.8096685 -18.724064826965332
        2.2686293 -21.29608154296875
      };
      \addplot[color=black,densely dotted] table {
        0. 0.41392685
        6. -35.70967263
      };
      \addplot[color=gray,densely dashed] table {
        0. -8.46706895786258
        6. -44.59066843786258
      };
      \addplot[color=black] table {
        0. 1.
        0. -10.
        6. -46.12359948
      };
      \legend{neural upper bound estimated with $L_\mathrm{m}$ (Fig.~\ref{fig:visualization_1}), neural upper bound estimated with $L_\mathrm{c}$, asymptotic R-D point-to-point, asymptotic R-D Wyner--Ziv + $1.53$ dB, asymptotic R-D Wyner--Ziv};
    \end{axis}
    \end{tikzpicture}
  \caption{$X=Y+N$ with $Y \sim \mathrm{N}(0,1)$ and $N \sim \mathrm{N}(0,10^{-1})$.}
  \label{fig:RD_var0.1}
\end{subfigure}%
\hfill%
\begin{subfigure}{\columnwidth}
    \raggedleft
    \begin{tikzpicture}[trim axis right]
    \begin{axis}[
      height=.19\textheight,
      width=.85\linewidth,
      scale only axis,
      xlabel={rate [bits]},
      ylabel={distortion [dB]},
      xmin=0.,
      xmax=3.,
      ymin=-25.,
      ymax=-10.,
      legend pos=north east,
      legend style={font=\scriptsize},
      ]
      \addplot[color=teal,mark=triangle*, mark size=2pt] table {
        1. -20.27697580250686
        2. -20.964133926201463
        3. -20.357155522665344
      };
     \addplot[color=blue,mark=square*] table {
        0.679996 -20.8489727973938
        0.89723724 -21.337780952453613
        1.2638619 -22.123868465423584
        1.4534979 -22.6043963432312
        1.8716173 -23.40317964553833
        2.3653505 -24.013419151306152
        2.7019472 -24.288642406463623
        3.2669206 -24.69569683074951
      };
      \addplot[color=red,mark=*] table {
        0.31479982 -21.115736961364746
        0.43293902 -21.552035808563232
        0.6946749 -22.629497051239014
        0.80841285 -23.079593181610107
        1.0567338 -24.226911067962646
        1.1426133 -24.82905149459839
      };
      \addplot[color=black,densely dotted] table {
        0. 0.0
        6. -36.12359947967774
      };
      \addplot[color=gray,densely dashed] table {
        0. -18.510282695689007
        6. -54.63388217536675
      };
      \addplot[color=black] table {
        0. 1.
        0. -20.043213737826427
        6. -56.16681321750417
      };
      \legend{neural distributed source coding~\cite{NDSC},
      neural upper bound estimated with $L_\mathrm{m}$,
      neural upper bound estimated with $L_\mathrm{c}$,  asymptotic R-D point-to-point, asymptotic R-D Wyner--Ziv + $1.53$ dB, asymptotic R-D Wyner--Ziv};
    \end{axis}
    \end{tikzpicture}
    \caption{$Y=X+N$ with $X \sim \mathrm{N}(0,1)$ and $N \sim \mathrm{N}(0,10^{-2})$.}
    \label{fig:RD_diff}
\end{subfigure}%
\caption{Rate--distortion (R-D) performances obtained with marginal and conditional formulations, as in Eqs.~\eqref{eq:proposed_loss_marginal} and \eqref{eq:proposed_loss_conditional}, respectively. We consider the quadratic-Gaussian WZ setup with two different correlation structures, and plot the empirical results versus the asymptotic bounds. The $1.53$ dB distortion offset refers to the space-filling loss that the entropy-constrained one-shot lattice quantizer is subjected to in a high-rate regime.}
\label{fig:RD}
\end{figure*}

We next consider the system model in Fig.~\ref{fig:sys_conditional}. The upper bound in Eq.~\eqref{eq:upper_bound_cond} corresponds to the rate of a system employing a one-shot quantizer and an ideal SW entropy coder which asymptotically achieves the cross-entropy $\mathbb{E}_{x}\big[\mathbb{E}_{u \sim p_{\boldsymbol{\theta}}(u \vert x)}[-\log q_{\boldsymbol{\zeta}}(u \vert y)] \big]$. Analogous to Section~\ref{sec:marginal}, minimizing $L_\mathrm{c}$ in Eq.~\eqref{eq:proposed_loss_conditional} corresponds to end-to-end optimization of this operational scheme.

The experimental results are provided in both panels of Fig.~\ref{fig:RD}. We observe that unlike the previous case, this model's performance is closer to the asymptotic WZ rate--distortion bound. We find no evidence of binning occurring in these quantizers (not depicted). We explain the improved rate--distortion performance of this model as follows. When binning is left to the ideal SW code, which may make use of a high dimensional channel code (e.g., as in DISCUS~\cite{DISCUS}), the performance loss of such a learned Wyner--Ziv compressor only comes from the quantization part alone. This line of reasoning was also followed by the practical code design in~\cite{tcq_ldpc}. The authors make use of a combination of a classic quantizer (without binning) and a powerful SW coding scheme, implemented with irregular low-density parity-check (LDPC) codes, in order to achieve the theoretical limit of $H(Q(X) \vert Y)$, where $Q(X)$ refers to the quantized source. Hence, minimizing $L_\mathrm{c}$ corresponds to learning one-shot quantizer and dequantizer components, reducing the WZ problem to a SW problem in a data-driven fashion.

\section{Discussion}
\label{sec:discussion}

We have proposed two solutions to the WZ problem, whose optimal theoretical solution is asymptotic and non-constructive. By establishing two variants of neural upper bounds, we have introduced constructive learning based-compressors, operating in the one-shot regime. Explicitly visualizing the behavior of these models, we provide post-hoc interpretations for the learned encoders and decoders. To ensure that the learning procedure cannot benefit from prior knowledge of the source, imposed into the design via special structure, we opted for a very generic parametrization of the probabilistic models. Fig.~\ref{fig:visualizations} provides the first explicit evidence of ANN-based learned compressors recovering some elements of the optimal theoretical solution, both through binning with respect to the source space, and piecewise linear behavior of the decoding function. Binning is a heavily used mathematical tool in information theory, and also characterizes practice-oriented schemes such as DISCUS~\cite{DISCUS}. Unlike the systematic partitioning of the quantized source space through cosets, as in DISCUS, our models are data-driven, and may find practical use for other sources beyond the Gaussian case, most of whose feasible solutions are unknown to this day. Our findings provide interesting data-driven insights about the nature of a classical source coding problem with side information.

In terms of constructive solutions, we have established the link between two neural upper bounds (Section~\ref{sec:method}) and two corresponding operational schemes (Sections~\ref{sec:marginal} and \ref{sec:conditional}) by picking a suitable entropy coding technique for each one. In the case of the marginal formulation in Eq.~\eqref{eq:proposed_loss_marginal}, it is attainable with high-order classic entropy coding, operating on discrete values. This choice is justified, as it has been shown that the actual rates achievable by a properly designed entropy code are only negligibly above the entropy values~\cite{universal_coding}. Considering the conditional formulation in Eq.~\eqref{eq:proposed_loss_conditional}, we make use of an ideal SW coding scheme~\cite{Slepian:IT:73}, which compresses sufficiently large blocks of quantized source elements to the rate of $H(Q(X) \vert Y)$. Unlike in the marginal variant, the operational role of SW coding is to additionally exploit the correlation between $Q(X)$ and $Y$ to yield further compression. This explains our empirical finding that in this case, there is no binning observed in the quantization (as SW coding takes care of this). State-of-the-art channel coding schemes such as LDPC~\cite{swc_nq, tcq_ldpc, ldpc_0, ldpc_3,  ldpc_4, girod_rate_adaptive_ldpc, girod_image_authentication} and turbo codes~\cite{turbo_0, turbo_3, turbo_4, turbo_1, turbo_2, blum_SW} have been demonstrated to yield results coming close to the theoretical SW bound. To be fair, in order to achieve optimality, these schemes make certain assumptions about the virtual channel, which might not be met in our case.

Previous work~\cite{NDSC} investigated the construction of neural WZ schemes using a machine learning technique called VQ-VAE~\cite{vq_vae}, which is comparable to the Concrete distribution formulation in that its objective is amenable to optimization using SGD. However, the objective does not explicitly model entropy, and instead contains a proxy objective that encourages utilization of all values of $u$. It thus does not correspond to an optimization of the rate--distortion Lagrangian (as in Eqs.~\eqref{eq:proposed_loss_marginal} and \ref{eq:proposed_loss_conditional}). The results of \cite{NDSC}, obtained for $Y=X+N$ with $X$ and $N$ being independent Gaussian, are reproduced in Fig.~\ref{fig:RD_diff}. We note that both of our methods outperform this scheme. We attribute the suboptimal performance of the scheme to the lack of explicit accounting for entropy in the learning objective.

Notable prior work on the machine learning side~\cite{VIB,CEB} is related to the \emph{information bottleneck} problem~\cite{tishby}. The learning objectives are comparable to our marginal and conditional formulation, respectively. However, both of these are strictly concerned with probabilistic model fitting, not with operational compression schemes. Reflecting this, model distributions are assumed Gaussian rather than discrete.

Going forward, by actually implementing the two aforementioned entropy coding techniques, and reporting the actual bit rates, we hope to demonstrate the feasibility of our neural schemes as a complete constructive end-to-end solution to the WZ problem. In the case of SW coding, learned channel coding techniques~\cite{kim_1, kim_2} could be investigated, to relax the assumptions about the virtual channel arising between the quantized source and side information.

\newpage

\IEEEtriggeratref{23}
\bibliographystyle{IEEEtran}

\bibliography{ref.bib}

\newpage

\end{document}